# Efficient Fuzzy Search Engine with B -Tree Search Mechanism

Simran Bijral
Department of Information Technology
Maharashtra Institute of Technology
Pune, India
simi.bijral1@gmail.com

Debajyoti Mukhopadhyay
Department of Information Technology
Maharashtra Institute of Technology
Pune, India
debajyoti.mukhopadhyay@gmail.com

*Abstract*— Search engines play a vital role in day to day life on internet. People use search engines to find content on internet. Cloud computing is the computing concept in which data is stored and accessed with the help of a third party server called as cloud. Data is not stored locally on our machines and the software's and information are provided to user if user demands for it. Search queries are the most important part in searching data on internet. A search query consists of one or more than one keywords. A search query is searched from the database for exact match, and the traditional searchable schemes do not tolerate minor typos and format inconsistencies, which happen quite frequently. This drawback makes the existing techniques unsuitable and they offer very low efficiency. In this paper, i will for the first time formulate the problem of effective fuzzy search by introducing tree search methodologies. I will explore the benefits of B trees in search mechanism and use them to have an efficient keyword search. I have taken into consideration the security analysis strictly so as to get a secure and privacy-preserving system.

## I. INTRODUCTION

There has been a rapid improvement of cloud computing, with a growing trend in cloud computing. Quite sensitive data is stored into the cloud such as personal health records, emails, personal chats, etc. Such data is to be encrypted before outsourcing into the cloud. There is a benefit of storing data into the cloud. When data is stored into the cloud, data owner does not have to worry about data storage and maintenance. Encryption of data has to be done properly because if not done properly, users will not retrieve data of their own interest. Users are more interested in data retrieval from the keyword search without paying any attention to the encrypted files back. But this kind of approach is not possible in cloud computing approaches. For example, Google Search, Bing Search, etc. uses this keyword search technique to retrieve data of interest to users.

Due to data encryption, users are not able to perform keyword search properly. Data encryption also asks for the protection of the keywords which makes it hard for users to perform keyword search.

In recent years, many solutions to this problem have been developed. [5-15]. There have been many researches on efficient and secure keyword search on Encryption schemes usually form an index table and trapdoor for keywords. encrypted data[8, 10, 11].This integration of the index table and trapdoor is an efficient method in preserving the privacy of the data. This helps in the preservation of both data and keyword. Traditional searchable techniques provide only exact keyword search. These techniques do not tolerate any kind of typos and format inconsistencies. This approach does not suit for cloud computing because users do not exactly know the keyword to be searched. Quite often, users' searching input does not exactly match the predefined set of keywords due to some typos and format inconsistencies, such as P.O Box can be written as P O Box.[2]User may not have the exact knowledge about the data.

This problem has been solved by proposing "Dictionary-based Fuzzy Set Construction"[3]in which a dictionary is maintained containing all legal English words. The keywords in index will all be legal English words. Stop words such as "a", "the", etc., will be excluded from the dictionary, because these words are never keywords. Another solution is the spell check algorithms, in which additional interaction of user is required to determine the exact word from the options generated by the spell check algorithms. One more disadvantage of this approach is that it cannot differentiate between two valid keywords, for example, if a user accidently types "mat" instead of "rat", this approach is not valid.

In this paper, we have formalized an effective but privacy-preserving fuzzy keyword in cloud computing. To the best of my knowledge, we have tried to maintain keyword privacy. We have used and explored the concepts of B tree for implementing the fuzzy keyword search. This kind of system will help in reducing the searching time and the privacy is maintained by encryption mechanism.

## II. RELATED WORK

### A. Predictions

With the help of this prediction technique [4], a system predicts a word which the user is going to type based on the partial input the user has given. There are some kind of queries that have multiple keywords as a single string. For example, if there is a search box on the home page of ICICIC Bank site, and we will search for "loans", then automatically it will show options like "Home loans". This kind of feature is called as autocomplete[4]. In this feature, two strings are more often used together rather than being used alone.

### B. Complete Search

Complete search provides user with exact match to their input query. In this, user types the query letter by letter, and the system finds records which are exactly similar to the query. If the query has multiple keywords, then Completesearch option will cache the result of the query without the last keyword. Incremental cache technique is often required for greater efficiency.

### C. Wildcard-Based Technique

The wildcard-based fuzzy set $w_i$ with edit distance ed is denoted as $Sw_i, ed=\{S'_{wi,0}, S'_{wi,1}, ...., S'_{wi,ed}\}$, where $S'_{wi,\tau}$ denotes the set of words $w'_i$ with $\tau$ wildcards[1]. Each wildcard represents an edit operation on $w_i$. For example, for the keyword STUDENT with the pre-set edit distance 1, its wildcard-based fuzzy keyword set can be produced as $S_{STUDENT, 1}$={STUDENT, *STUDENT, *TUDENT, S*TUDENT, S*UDENT, ..., STUDEN*T, STUDEN*, STUDENT*}.

## III. PROPOSED WORK

### A. Dictionary-based Search

A dictionary is a set of all possible English words. Let $e_1, e_2,......,e_n$ are n distinct words and their set is D= $\{e_1, e_2,......, e_n\}$. All the fuzzy keywords should belong to the dictionary. For example, I can make use of English dictionary which contain all the legal English words except some stop words because these stop words such as "a", "the", etc., are not included in keywords.

### B. String Edit Distance

There are three kind of operations that can be performed on strings:
(1)Insertion: We can insert one character anywhere in the string.
(2)Deletion: We can delete one character from the string.
(3)Substitution: We can replace a character in the string.

String edit distance is the minimum number of operations performed on a given strings to transform it into another string.

### C. Fuzzy search

Given a keyword w and a dictionary D, let $F^D_{w,d}$= $\{w_1,w_2,w_3,....,w_q\}$ be the fuzzy keywords of word w, whose string distance is d, for all $1\leq i\leq q$, wi belongs to D and $0\leq ED(w,w_i)\leq d$.

For example, if we use a dictionary $D_0$ containing all legal English words, for example , if the keyword is "computer" and the string edit distance is 1, $F_{w,d}{}^D{}_0$={computer, compute, commuted, computes, computers, commute}

As in cloud computing, we often face problems related to data security, I have tried to construct a way in which users' data will be secure and less susceptible to malicious attack. By building an index table, data will be less prone to attacks. Index table will contain trapdoors and encrypted data. $T_{wi}$ denote a trapdoor of keyword $w_i$. It can be calculated as $T_{wi}= f(sk,w'_i)$ for each $w'_i$ $S_{wi,d}$ where $S_{w,d}$ denote the set of all words $w_i$ satisfying $ED(w,w_i)\leq d$.

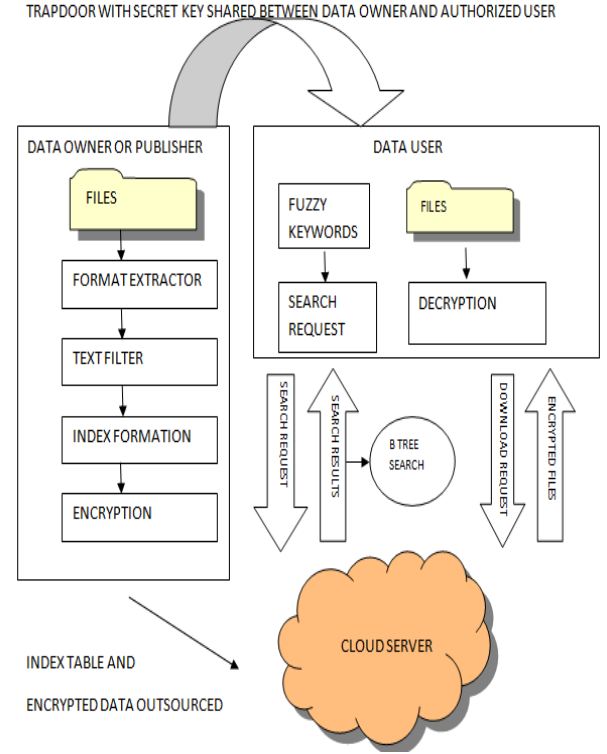

Figure 1. Proposed Architecture

The data owner computes the trapdoor and shares it with the authorized users with a secret key sk. Secret key for authorized user will be nothing but username and password.

Now, when a user search a keyword w, then the authorized user will compute the trapdoor $T_w$ for w and will send it to the server. Now, after receiving $T_w$ from user, server will compare it with the index table and will return all possible encrypted files identifiers {Enc(sk,$FID_{wi}$)}, where $FID_{wi}$ is the set of all possible Ids whose corresponding data files contain the word $w_i$. Suppose, if the input keyword is (w,k) with edit distance k≤ d, then the system will return set of all possible Ids whose corresponding data files contain the word w. It is denoted as $FID_w$, if w=$w_i$ and $w_i$   W, then the system will return $FID_{wi}$, or else, if w does not belong to W, then the system will return {$FID_{wi}$}

### D. Different sections

*Different sections in Data owner:*
As shown in the Figure 1, there are different modules in the architecture, they are:
Files: there are set of files which are ready to be outsourced to the cloud server by the data owner.
Format Extractor: different files have variety of formats, such as doc file, txt and xls. It will read the format and extract the content of the file.
Text Filter: Other than the legal English words in the text there are some other information such as, punctuation, separators, etc. It is used to filter these kind of information and helps in generating a list of words.
Index Formation: Based on the above explained architecture, i.e, Dictionary-based Fuzzy keyword search, we will form an index table for each keyword as explained in the above paragraphs.
Encryption: I have used MD5 encryption algorithm. The data owner will share the secret key to authorized data user via login details.

*Different sections in Data user*:
Fuzzy Keywords: With the input keyword w, the data user will generate the fuzzy keyword set of w with some string edit distance.

### E. B Tree search Mechanism

B-Tree is a data structure that stores the data and allows various operations, such as, searches, insertion, deletion, etc. It is a generalization of binary search tree in which there can be more than two children. A B Tree of order m has root node with at least two child nodes and at most m child nodes. The internal nodes except the root have at least ⌈m/2⌉ child nodes at most m child nodes. The number of keys in each internal node is one less than the number of child nodes. All leaf nodes are on the same level.

*Pseudo Code*
2 inputs: x, pointer to the root node of a subtree, k, a key to be searched in that subtree.
Function B-TREE-SEARCH(x,k) returns(y,i) such that $key_i$[y]=k or NIL
  i←1
while i ≤ n[x] and k > $key_i$[x]
  do i ← i + 1
    if i ≤ n[x] and k = $key_i$[x]
then return (x,i)
if leaf[x]
    then return NIL
ELSE disk-read($c_i$[x])
    return B-TREE-SEARCH($C_i$[x], k)

At each internal node x we make an
(n[x] +1)- way branching decision.
Indexing of data is also used along with B Tree search. Inverted index is used to map words to its corresponding document numbers(s). Fully inverted index holds the word's position in the corresponding document number(s). For example, we have 3 documents $D_0$, $D_1$ and $D_2$ as follows:
$D_0$= "Everyone likes Aishwarya_Rai"
$D_1$= "Aishwarya_Rai is a Bollywood actress"
D2= "There is no one with the likes of Aishwarya_Rai

Table 1. Inverted Index Table

| Inverted Index | | Fully Inverted Index |
|---|---|---|
| "Everyone" | {0} | {(0,0)} |
| "likes" | {0,2} | {(0,1), (2,6)} |
| "Aishwarya_Rai" | {0,1,2} | {(0,2), (1,0), (2,8)} |
| "is" | {1,2} | {(1,1), (2,1)} |
| "a" | {1} | {(1,2)} |
| "Bollywood" | {1} | {(1,3)} |
| "actress" | {1} | {(1,4)} |
| "There" | {2} | {(2,0)} |
| "no" | {2} | {(2,2)} |
| "one" | {2} | {(2,3)} |
| "with" | {2} | {(2,4)} |
| "the" | {2} | {(2,5)} |

Now, when the user will search for the keyword "Everyone", the index table is searched and the document where this search string is found, is displayed. When there is a multiple keyword search, then we take the join of all keywords and display the result[9]. If, user searches for the keywords "likes" and "Aishwarya_Rai", a search from inverted index is carried out and it computes {0,2}   {0,1,2}={0,2}. This will list the number of documents in which the search string is found. A search on the fully inverted index will result in:

"likes" →{(0,1), (2,6)}

"Aishwarya_Rai"→{(0,2), (1,0), (2,8)}

This means "likes " is the word in position 1 in $D_0$ and is in position 6 in $D_2$, and "Aishwarya_Rai" is the word in position 2 in $D_0$, the word is in position 0 in $D_1$, the word is in position 8 in $D_2$, starting from position 0.

Searching in B Tree is done normally and the result is searched in inverted index table for the document to be displayed. Searching in B Trees is similar to that of binary search trees. For example the given tree is:

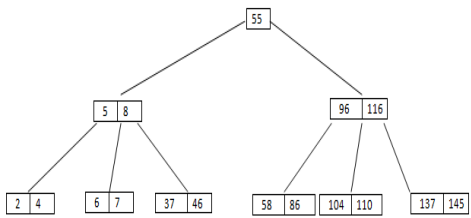

Figure 2. B Tree

To search for 37, we will begin from the root and as 37 < 55 we will look on the right hand side. 37>8, so we will go on the left side of 8. We will get 37 there. Similarly, character searching is performed by assigning numerals like 1, 2, ..., 26 to a, b, ..., z. Once we get the alphabet, we store it in a buffer and search for another one. Once the word is completed, then we look for that word in the inverted index table and visit the required document.

IV. IMPLEMENTATION SNAPSHOTS

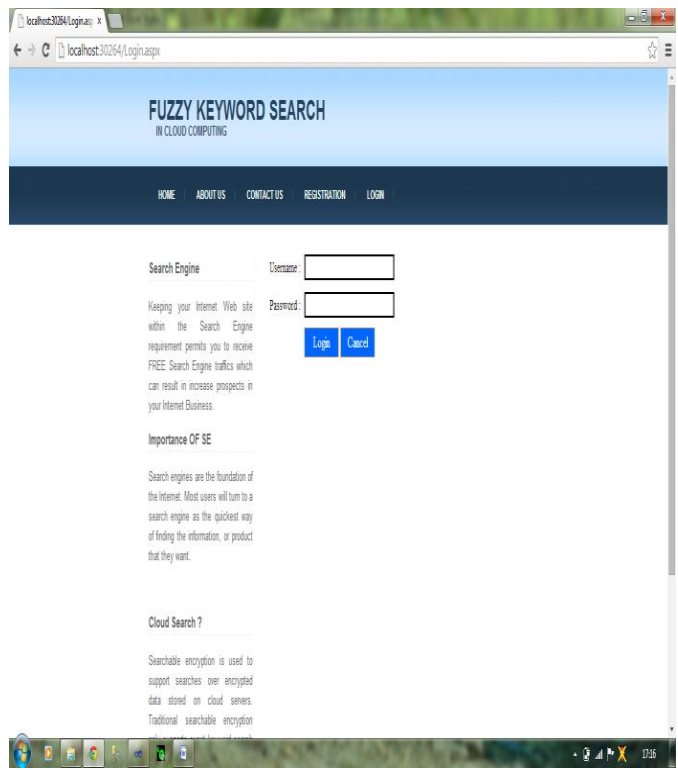

Figure 3. Screen Snapshot of Login Page

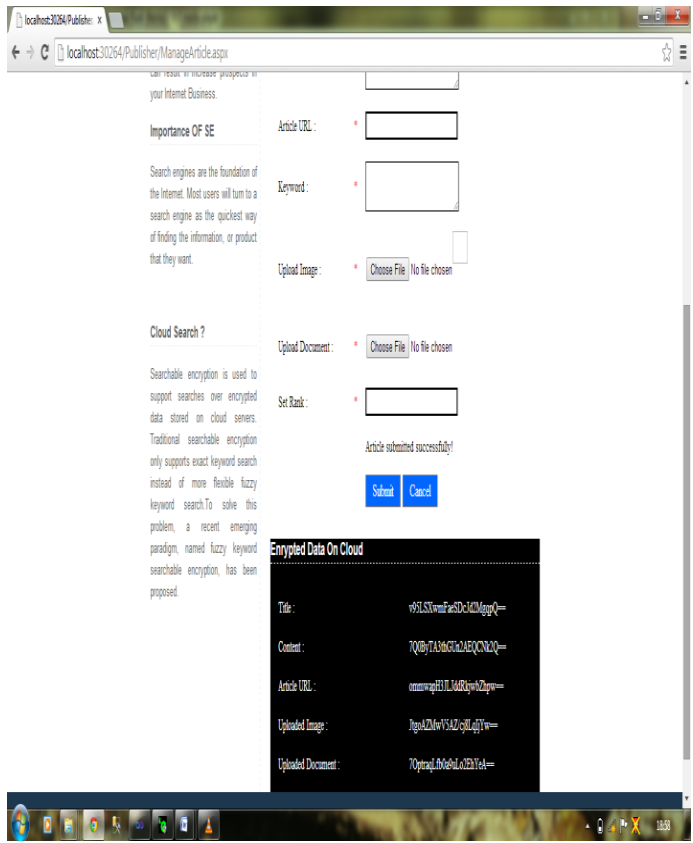

Figure 4. Screen Snapshot of Publisher Module

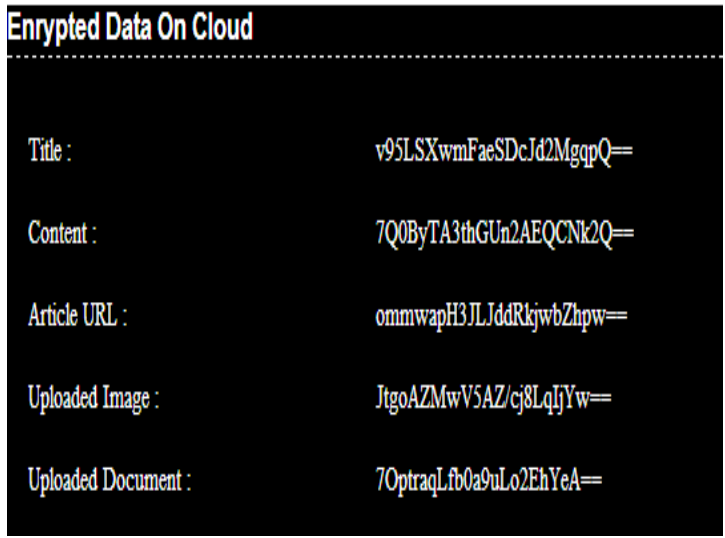

Figure 5. Screen Snapshot of Encryption of Data

This Figure 5 shows the encryption of the data that the publisher stores on the cloud.

## V. COMPARISON BETWEEN DICTIONARY-BASED AND WILDCARD-BASED TECHNIQUE

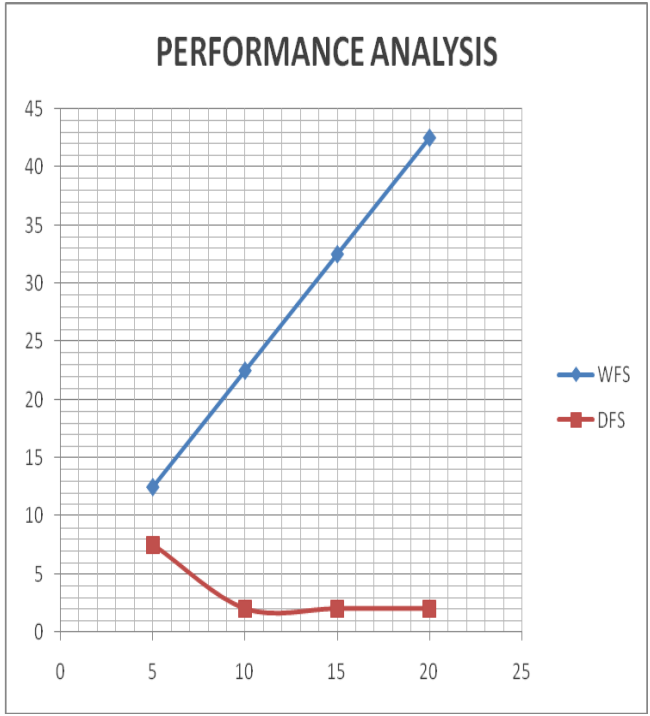

Figure 6. Performance Analysis

WFS is Wildcard-Based Search
DFS is Dictionary-Based Search
x-axis represents Keyword length

y-axis represents average number of fuzzy keywords Edit-Distance used here is 1.

The Figure 6 shows that fuzzy keywords in WFS are much more than the fuzzy keywords in DFS, so, DFS has smaller search request size. If the word to be searched is longer, then the DFS will give lesser number of fuzzy keywords. For example, if the keyword is STUDENT, then using Wildcard-Based Technique, the number of fuzzy words will be much larger than the words using Dictionary-Based Search Technique. STUDENT keyword will give words like ASTUDENT, BSTUDENT, ….., ZSTUDENT, *TUDENT, S*TUDENT, ST*UDENT, and so on, using WFS, which is meaningless because these are not valid english words. So, for this problem DFS is better technique.

## VI. CONCLUSION

In this paper, we presented a complete architecture of fuzzy keyword search scheme. Through the results, we suggest that Dictionary-based Fuzzy Keyword Search is an efficient method to make fuzzy keyword search scheme. Time taken by Dictionary–based fuzzy keyword search is much less than the time taken by any other method. For searching the keyword, we have explored the benefits of B Trees and our proposed system is secure and privacy- preserving, therefore, it correctly realize the function of Fuzzy keyword search. This solution has features like:

1. Text to speech search is there, so, the user can recheck the data which he has entered,
2. History of the user is provided, so that he can check the history.

This searching technique works well on every data set such as database table entry, data matrix, column, etc.

As our ongoing work, we will continue to research on the security related to the cloud that will support:
1. search that will also take conjunctions of words,
2. Ranking of keywords according to user preferences.

### REFERENCES

[1] B. Waters, D. Balfanz, G. Durfee, and D. Smetters, "Building an encrypted and searchable audit log", Proceeding of 11th Annual network and Distributed System, 2004.
[2] Chang Liu, Liehuang Zhu, Longyijia Li and Yu'an Tan, "Fuzzy Keyword Search on Encrypted Cloud Storage Data with Small Index", Proc. Of IEEE CCIS'11, pp. 269-273, 2011.
[3] D. Boneh, G. D. Crescenzo, R. Ostrovsky, and G. Persiano, "Public key encryption with keyword search", Proceedings of EUROCRYP,04, 2004.
[4] D.Song, D. Wager and A. Perrig, "Practical techniques for searches on encrypted data", Proceedings of IEEE Security and Privacy, 2000.
[5] E. J. Goh, "Secure Indexes", Cryptography ePrint Archive, Report 2003/216, 2003.
[6] F. Bao, R. Deng, X .Ding and Y. Yang "Private query on encrypted data in multi-user settings", in Proceedings of ISPEC'08, 2008.
[7] Jin Li, Qian Wang, Cong Wang, Ning Cao, Kui Ren and Wenjing Lou, "Fuzzy Keyword Search over Encrypted Data in Cloud Computing", Proc. IEEE INFOCOM, 2010.
[8] J.W.Byun, D.H.Lee, and J. Lim, "Efficient Conjunctive Keyword Search on Encrypted Data Storage System," Lecture Notes in Computer Science, Volume 4043, Public Key Infrastructure, Pages 184-196, 2006.
[9] M.Chuah and W.Hu, "Privacy-aware BedTree Based Solution for Fuzzy Multi-Keyword Search over Encrypted Data", 31st International Conference on Distributed Computing Systems Workshops, pp. 273-281, 2011.
[10] M. Abdalla, M. Bellare, D.Catalona, E.Kiltz, T.Kohno, T.Lange, J.Malone-Lee, G.Neven, and P.Paillier, "Searchable Encryption Revisited: Consistency Properties, Relation to anonymous IBE, and extensions," Journal of Cryptology, Volume 21, Number3, Pages 350-391, 2008.
[11] M. Bellare, A. Boldyreva, and A. O'Neil, "Deterministic and Efficiently Searchable Encryption," in Proc. of Crypto 2007, Volume 4622 of LNCS. Springer-Verlag, 2007.
[12] Ranjeeth Kumar.M and D.Vasumathi, "A Novel Implementation Of Fuzzy Keyword Search over Encrypted Data in Cloud Computing", International Journal of Computer Trends and Technology, pp.275-279, July-Aug 2011.
[13] R. Cutmola, J.A.Garay, S.Kamara, R.Ostrovsky, "Searchable symmetric encryption: improved definition and efficient constructions", Proceedings of ACM CCS, 2006.
[14] Shengyue Ji, Guoliang Li, Chen Li and Jianhua Feng, "Efficient Interactive Fuzzy Keyword Search", International World Wide Web Conference Committee(IW3C2), pp. 371-380, 2009.
[15] Y-C Chang and M. Mitzenmacher, "Privacy preserving keyword searches on remote encrypted data", Proceedings of ACNS'05, 2005.